\documentstyle[11pt,newpasp,twoside,epsf]{article}
\def\edcomment#1{\iffalse\marginpar{\raggedright\sl#1\/}\else\relax\fi}
\reversemarginpar

\begin{document}
\title{21st Century VLBI: Deep Wide-Field Surveys}

\author{M.~A. Garrett}
\affil{Joint Institute for VLBI in Europe, Dwingeloo, NL}
\author{J.~M. Wrobel}
\affil{National Radio Astronomy Observatory, Socorro, NM, U.S.A.}
\author{R. Morganti}
\affil{Netherlands Foundation for Radio Astronomy, Dwingeloo, NL}

\begin{abstract}
A 20-hour VLBI observation with the NRAO VLBA and GBT in the NOAO
Bo\"otes field reaches an rms noise of 9 microJy per beam at 1.4~GHz.
Three sources were detected at 10-milliarcsecond resolution within the
GBT primary beam of FWHM 8.6\arcmin, including the 20-milliJy
calibrator and two sub-milliJy sources.  By tapering the visibility
data, portions of the VLBA primary beam of FWHM 29\arcmin\, were
imaged at poorer sensitivity and resolution to yield five further
detections.  New developments at JIVE will permit deeper and wider
VLBI surveys at full sensitivity and resolution, enabling new types of
survey science.
\end{abstract}

\section{Motivation}

Very Long Baseline Interferometry (VLBI) observations, and VLBI
surveys in particular, have contributed significantly to our
understanding of active galaxies.  For example, the discovery of
superluminal motion in these systems provided the first clue that
viewing angle was an important parameter in interpreting and
eventually unifying different classes of active galaxies, including
radio galaxies and quasars (Urry \& Padovani 1995).  More recently,
VLBI has provided direct evidence for an evolutionary scenario wherein
very young compact sources are the precursors of giant extended
sources associated with radio galaxies (Taylor et al.\ 2000).
Numerous other advances are summarized in these Proceedings.

At GHz frequencies, VLBI surveys usually target sources distributed
across the sky.  The sources are also often pre-selected to be strong
($S_{T} > 200$~mJy) and to exhibit flat spectra (e.g., Beasley et al.\
2002).  Such samples are largely dominated by intrinsically luminous
active galaxies at moderate redshifts $z\sim 1-2$.  Self-calibration
techniques can be used for such bright sources so calibration overhead
is minimal.  But the observation time $t$ per source is reduced by
needing to slew the array from source to source.  Also, most observers
impose a typical image field of view ($fov$) of $\sim
100$~milliarcseconds, ensuring that only one source is detected in any
given observation time $t$.  Such VLBI surveys are therefore slow and
the biases introduced are substantial.

Recent attempts have been made to survey weaker ($S_{T} > 10$~mJy)
sources using phase referencing in the switching style.  Such VLBI
surveys take advantage of the huge catalog of FIRST sources now
available at 1.4~GHz with a resolution of 5\arcsec\, (White et al.\
1997).  These VLBI surveys are usually localized to a region
subtending a few square degrees and including a strong phase
calibrator close to weaker targets.  Example surveys are presented in
these Proceedings by Gurvits et al.\ (2004) and by Wrobel et al.\
(2004).  These surveys are both unbiased as no spectral pre-selection
is made and efficient as they minimize array slewing.  But phase
referencing in the switching style causes substantial calibration
overhead, leaving a typical observation time per source of $t \sim
10$~minutes.  For the weaker VLBI detections, the dynamic range can be
10 or less.  For the stronger VLBI detections, the dynamic range is
often limited by residual phase-referencing errors.  These limitations
mean that jets and other secondary features can easily be missed,
making morphological classification difficult.  Although a $fov \sim
1000$~milliarcseconds is required to accommodate FIRST's astrometric
accuracy, this is still small enough to ensure only one target per
$fov$.  Moreover, the 10-milliJy sources from FIRST still involve
essentially the same radio source populations that are targeted in the
bright surveys (Ivezi\'c et al.\ 2002).

Indeed, at 1.4~GHz most sources slightly stronger than $S_{T} \sim
1$~mJy are energized by active nuclei (e.g., Condon et al.\ 2003).
What steps can be taken toward studying the more diverse sub-milliJy
population at resolutions of, say, 10~milliarcseconds (80~pc at $z
\sim 2$ for $\Omega_m = 0.3$, $\Omega_{\Lambda} = 0.7$, and $H_0 =
70$~km~s$^{-1}$~Mpc$^{-1}$)?  Sources brighter than 0.1~milliJy have
an areal density of 0.2 per square arcminute, so about a dozen are
expected within the primary beam of a 100-m class antenna.  Therefore,
progress can be expected if two technical goals can be reached:
achieve wider fields of view and lower detection thresholds, such that
many sources can be observed simultaneously with the VLBI array.

For a connected array with short baselines, the $fov$ is often set by
the primary-beam response of the antennas in the array.  For VLBI this
is rarely the case.  In VLBI, a more demanding limitation is set by
the frequency and time sampling used during data correlation.  This
sampling must be fine enough to mitigate the effects of both bandwidth
and time-average smearing in images.  Also, since preserving $fov$
scales computationally with the square of the baseline length,
generating wide-field VLBI images is often limited by the off-line
computing resources available.  This latter restriction has introduced
a psychological barrier, which is that most VLBI observers
over-average their data in both the time and frequency domains to make
data sets more manageable.  Data-averaging collapses both the $fov$
and the total information content, leaving us with the postage-stamp
VLBI images with which we are all so familiar.

\section{Deep, Wide-Field EVN Imaging in the HDF-N}

The sub-milliJy population in a region containing the Hubble Deep
Field-North (HDF-N) was observed using the EVN at 1.6~GHz,
25-milliarcsecond resolution, and 64-MHz bandwidth (Garrett et al.\
2001).  The observations spanned two 16-hour segments but phase
referencing in the switching style left an observation time of only $t
= 14$~hours.  A $fov$ a few arcminutes in diameter was set by
bandwidth and time-average smearing, and by the primary-beam responses
of the largest antennas in the array.  The EVN clearly detected
sources stronger than 210~microJy per beam (5$\sigma$) in two of the
five target patches imaged within the $fov$.  One detection locates
the active nucleus and an adjacent feature in an FR-I galaxy at $z =
1.0$, while the other detection is a candidate active nucleus embedded
in an obscured starburst at $z = 4.4$.  The observed rms noise levels
were higher than the expected thermal levels, due to residual
phase-calibration errors.

\section{Deep, Wide-Field VLBA+GBT Imaging in Bo\"otes}

We have improved upon the HDF-N survey with a VLBI observation in the
NOAO Bo\"otes field using the NRAO VLBA and GBT at 1.4~GHz and 64-MHz
bandwidth (Garrett, Wrobel, \& Morganti, in preparation).  The VLBA is
described by Napier et al.\ (1994).  The observations spanned three
8-hour segments.  Phase referencing in the in-beam style led to an
observation time $t = 20$~hours.  Two correlation passes were made
within the GBT primary beam of FWHM 8.6\arcmin, shown as the circle in
the figure.  The passes were centered on the in-beam calibrator and on
a region with six faint WSRT targets (de Vries et al.\ 2002, Morganti
\& Garrett 2002).  For each pass, a $fov \sim$ 4\arcmin\, in diameter
was preserved and the output data rate of 1~MByte per second yielded a
data set of size 60~GBytes.

The data from the calibrator pass were edited, averaged, and
calibrated, and its gain parameters applied to the target-pass data.
The AIPS task IMAGR was used to make dirty images/beams from the
target-pass data after editing and calibration.  Each dirty image
subtended a square 6\arcsec\, on a side.  For each target, three such
dirty images were made from the three observation segments and then
simply co-added.  This was a slow computational task: about 8 hours
were required to produce one dirty image/beam on a 2-GHz Linux PC with
dual processors.  For the stronger sources, cleaned images were
produced with AIPS task APCLN by subtracting the dirty beam from the
dirty image.  Visibility-based cleaning is currently prohibitively
expensive in terms of computing requirements.

Three sources were detected within the GBT primary beam, at a
resolution of 10~milliarcseconds and above the detection threshold of
63~microJy per beam (7$\sigma$).  These detections include the
20-milliJy calibrator and two sub-milliJy sources.  Each detection is
unresolved and has NOAO photometric data available (Januzzi \& Dey
1999).  The calibrator is a quasar.  One sub-milliJy detection locates
the active nucleus with $S_{VLBI} \sim 0.4$~milliJy within an FR-I
galaxy with $S_{T} \sim 6.8$~milliJy.  The other sub-milliJy detection
has $S_{VLBI} \sim 0.4$~milliJy and recovers most of the $S_{T} \sim
0.6$~milliJy.  This is a candidate active nucleus because, at the
comological distance suggested by the host's blue drop-out, (1) the
spectral luminosity is too high for prompt emission from a SN and (2)
the WSRT photometry spanning 20 months shows no variability, arguing
against emission from a GRB afterglow.  Finally, the disk galaxy
NGC\,5646 at $z \sim 0.03$ with $S_{T} \sim 2.7$~milliJy was not
detected at 10-milliarcsecond resolution; its WSRT emission follows
the optical isophotes and obeys the FIR/radio correlation, suggesting
a star formation origin.

By tapering the visibility data, portions of the VLBA primary beam of
FWHM 29\arcmin\, were imaged at poorer sensitivity and resolution to
yield five further detections.  These detections involve brighter
radio sources probably hosted by active galaxies and quasars, but the
available NOAO photometric data have not yet been analysed.

The long-term goal for these new VLBI surveys is to constrain the
contribution that active galaxies make to the faint radio source
population in general, and the optically faint radio/sub-mm source
population in particular (Garrett 2003).  To have an impact in this
area, it is essential that many sources be surveyed over large regions
with rms sensitivities of a few microJy.  Global VLBI arrays employing
disk-based recordings should be able to achieve such sensitivities.
The proposed PCInt program at JIVE will enable output data rates as
high as 160~MBytes per second, remarkably expanding the accessible
$fov$.  It will thus be possible to sample simultaneously the summed
response of all milliarcsecond-scale emission within (and beyond) the
primary-beam response.  After some phase-referencing to improve the
coherence time, simple self-calibration of most fields should be
possible at frequencies of 1-3~GHz.  Access to GRID-like computing
resources may be the best way to analyse the enormous data sets
generated.  Huge and unbiased VLBI surveys will be conducted, with the
bulk of the targets being faint sources with flux densities of only a
few tens of microJy.

\acknowledgments This work made use of images provided by the NDWFS,
which is supported by NOAO.  NOAO is operated by AURA, Inc., under a
cooperative agreement with the NSF.  NRAO is a facility of the NSF
operated under cooperative agreement by AUI.

\begin{figure}
\plotone{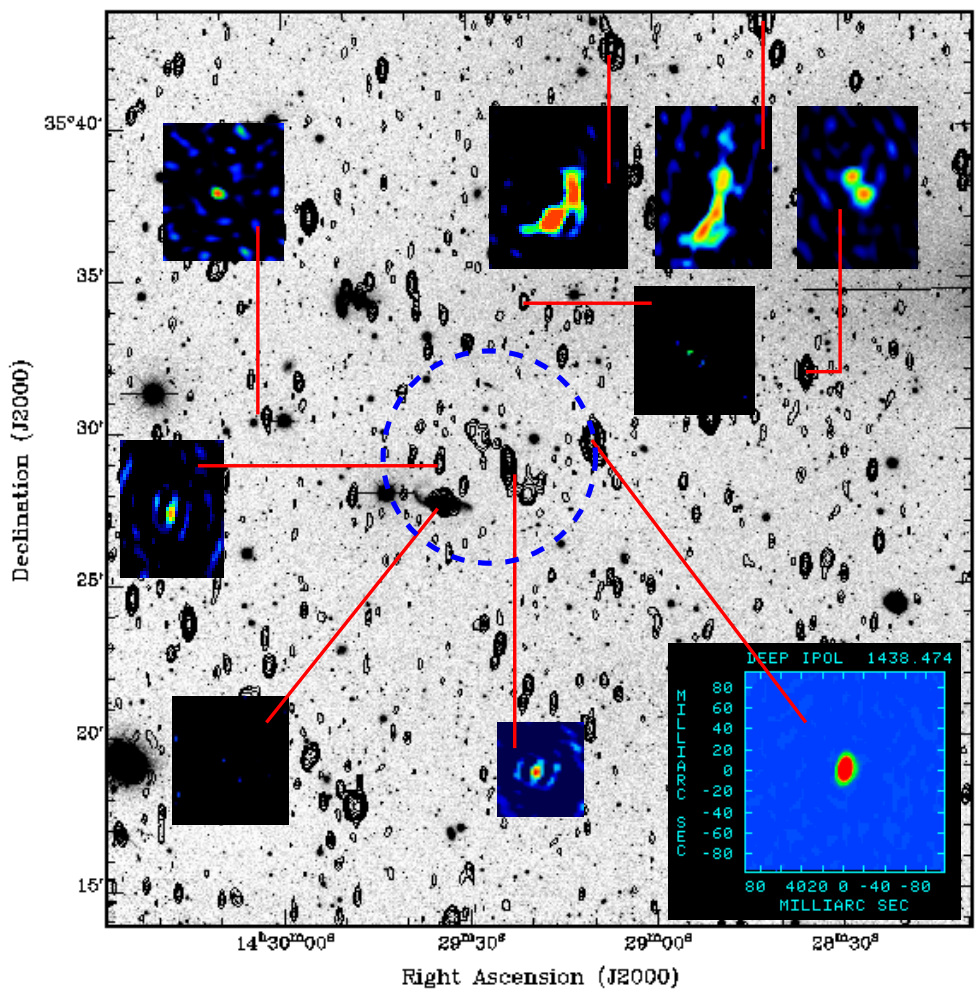}

\caption{Stokes I emission at 1.4~GHz (contours) for a portion of the
  NOAO Bo\"otes field (grey scale) imaged with the WSRT with a
  detection threshold of 65~microJy per beam ($5\sigma$, Morganti \&
  Garrett 2002).  Insets show Stokes I emission at 1.4~GHz from a
  20-hour VLBI observation with the VLBA plus GBT.  Within the circled
  region, the VLBI detection threshold is 63~microJy per beam
  ($7\sigma$) and the resolution is 10 milliarcseconds.  One inset
  shows a VLBI non-detection of a nearby star-forming disk galaxy
  easily detected by the WSRT.  Beyond the circled region, the VLBI
  detection threshold is higher and the resolution is poorer because
  the visibility data were tapered.}
\end{figure}

\end{document}